\documentclass[preprint,aps,superscriptaddress,nofootinbib]{revtex4}
\usepackage{amssymb}
\usepackage{amsmath}
\usepackage{graphicx}
\usepackage{color}

\newcommand\beq{ \begin{eqnarray} }
\newcommand\eeq{ \end{eqnarray} }

\begin{document}

\title{A Holographic Model for Hall Viscosity}
\author{Jiunn-Wei Chen\thanks{%
E-mail: jwc@phys.ntu.edu.tw}}
\affiliation{Department of Physics and Center for Theoretical Sciences, National Taiwan
University, Taipei 10617, Taiwan}
\affiliation{Leung Center for Cosmology and Particle Astrophysics\\
National Taiwan University, Taipei 106, Taiwan}
\author{Nien-En Lee\thanks{%
E-mail: r99222006@ntu.edu.tw}}
\affiliation{Department of Physics and Center for Theoretical Sciences, National Taiwan
University, Taipei 10617, Taiwan}
\author{Debaprasad Maity\thanks{%
E-mail: debu.imsc@gmail.com}}
\affiliation{Department of Physics and Center for Theoretical Sciences, National Taiwan
University, Taipei 10617, Taiwan}
\affiliation{Leung Center for Cosmology and Particle Astrophysics\\
National Taiwan University, Taipei 106, Taiwan}
\author{Wen-Yu Wen\thanks{%
E-mail: steve.wen@gmail.com}}
\affiliation{Department of Physics, Chung Yuan Christian University, Chung Li City, Taiwan}
\affiliation{Leung Center for Cosmology and Particle Astrophysics\\
National Taiwan University, Taipei 106, Taiwan}

\begin{abstract}
We have modified the holographic model of Saremi and~Son \cite{Saremi:2011ab}
by using a charged black brane, instead of a neutral one, such that when the
bulk pseudo scalar ($\theta $) potential is made of $\theta ^{2}$ and $%
\theta ^{4}$ terms, parity can still be broken spontaneously in the boundary
theory. In our model, the 3+1 dimensional bulk has a pseudo scalar coupled
to the gravitational Chern-Simons term in the anti de Sitter charged black
brane back ground. Parity could be broken spontaneously in the bulk by the
pseudo scalar hairy solution and give rise to non-zero Hall viscosity at the
boundary theory.
\end{abstract}

\maketitle



\section{Introduction}

In recent years, the AdS/CFT correspondence \cite{Maldacena,Polyakov,Witten}
has been applied to study strongly coupled phenomena in condensed matter
physics at finite temperature and chemical potential. In particular,
inspired by the idea of spontaneous symmetry breaking in the presence of
horizon \cite{Gubser:2005ih,GubserPRD78}, holographic superconductors \cite%
{Hartnoll, Horowitz} and superfluids \cite{Herzog:2008he} are two remarkable
examples where the Gauge/Gravity duality plays an important role.

On the other hand, the hydrodynamic limit of AdS/CFT correspondence has also
attracted much attention recently. Computations of the ratio of shear
viscosity to entropy density for a big class of gauge field theories with
gravitational duals yields the same number $1/4\pi $ which is not far away
from that observed in the strong interacting quark-gluon plasma created in
RHIC \cite{Kovtun:2004de,Son:2007vk}. Later it has been shown that by using
the boundary derivative expansion, one can consistently solve the Einstein
equation order by order and compute various hydrodynamics transport
coefficients of the boundary fluid \cite{minwalla}. Recently, a holographic
model for the parity violating Hall viscosity was proposed. Like the other
transport coefficients, Hall viscosity is also found to be uniquely
determined by the near horizon data of the bulk black brane \cite%
{Saremi:2011ab}. This is yet another example of the membrane paradigm. In
the original construction the $(3+1)$ dimensional bulk action has a negative
cosmological constant, a real scalar field coupled to the gravitational
Chern-Simons term\footnote{%
There exist early studies of Chern-Simons term in the Holographic models. To
mention a few: the effect of Maxwell Chern-Simons term $\theta {}\ ^{\ast
}FF $ was studied in the Holographic superconductor\cite%
{Tallarita:2010vu,Tallarita:2010uh}. The spectrum of quasinormal modes was
studied in the dynamic Chern-Simons gravity and correction to some
hydrodynamic quantities was discussed\cite{Delsate:2011qp}.}.

While it has been shown that a non-trivial profile of the bulk scalar field
is important to obtain a nonvanishing Hall viscosity of the $(2+1)$
dimensional boundary field theory, from the holography point of view it
would be interesting to further investigate what role this bulk scalar plays
at the boundary. One possible interpretation is to identify the boundary
value of this scalar as an order parameter field which condensates at low
temperature in the boundary field theory. From the condensed matter point of
view the physical realization of this order parameter, which leads a system
to the spontaneously parity breaking phase is not clear. But interestingly
in terms of physical quantity such as hall viscosity one might get
information about how the system breaks parity spontaneously. So,
effectively in the hydrodynamic regime, Hall viscosity can play the role of
order parameter which is non-zero only below a critical temperature. To be
ready for such a boundary theory interpretation, one shall look for a
sourceless boundary condition for the hairy scalar\textbf{\ }if parity is
only broken spontaneously.

However, it has been shown that a neutral scalar hair with quadratic and
quartic potential that satisfies the usual sourceless boundary condition in
a Schwarzschild-AdS black hole spacetime does not satisfy the positive
energy theorem \cite{Hertog:2006rr}. This essentially means that a
Schwarzschild-AdS black hole with a sourceless neutral scalar hair is
intrinsically unstable.

While it is still possible to find a sourced solution which minimizes the
free energy, we will take a different approach to modification of the model
by including a gauge field in the bulk. The scalar in the original theory is
identified as a pseudo scalar now, so its coupling to the gravitational
Chern-Simons term does not break parity. The pseudo scalar hair, however,
breaks parity spontaneously and gives a pseudo scalar condensate in the
boundary field theory which, as we will demonstrate in the next section, is
important for Hall viscosity. In the probed limit, this pseudo scalar hair
solution in the charged black brane background is known to be stable \cite%
{Iqbal:2010eh}.

The paper is organized as follows: in section two, we present a general
discussion of the parity violating viscosities and set up the holographic
model. We then compute the Hall viscosity and commend on the boundary field
theory in section three. We then conclude our results in section four. A
detailed derivation of Hall viscosity together with an analytical
approximation are given in the appendix.

\section{General properties of viscosities}

It is instructive to classify viscosities by considering the general
relation between the energy momentum tensor and the spatial derivative of
the fluid velocity%
\begin{equation}
T_{ij}=\eta _{ijkl}\partial _{(k}V_{l)}+\xi _{ijkl}\partial _{\lbrack
k}V_{l]},
\end{equation}%
where $i$, $j$,$k$,$l$ are spatial indices and $\partial _{(k}V_{l)}=\left(
\partial _{k}V_{l}+\partial _{l}V_{k}\right) $ and $\partial _{\lbrack
k}V_{l]}=\left( \partial _{k}V_{l}-\partial _{l}V_{k}\right) $ are just the
symmetric and anti-symmetric combinations of the derivatives, respectively.
We have $T_{ij}=T_{ji}$. In two spatial dimensional systems, $\eta _{ijkl}$
and $\xi _{ijkl}$ can be constructed by $\delta _{ij}$ and the
two-dimensional antisymmetric tensor $\epsilon _{ij}$. Taking $\eta
_{ijkl}\propto \delta _{ij}\delta _{kl}$, $\delta _{ik}\delta _{jl}+\delta
_{jk}\delta _{il}$ and $\epsilon _{ik}\epsilon _{jl\ }+\epsilon
_{jk}\epsilon _{il}$\ give rise to the usual shear ($\eta $) and bulk ($%
\zeta $) viscosity contributions 
\begin{equation}
\delta T_{ij}=-\eta (\partial _{i}V_{j}+\partial _{j}V_{i}-\text{trace}%
)+\zeta \delta _{ij}\mathbf{\nabla }\cdot \mathbf{V}.
\end{equation}%
Taking $\eta _{ijkl}\propto \delta _{ik}\epsilon _{jl\ }+\delta
_{jk}\epsilon _{il}$ gives rise to the Hall viscosity ($\eta _{A}$) and
\textquotedblleft curl\textquotedblright\ viscosity ($\zeta _{A}$)
contributions%
\begin{equation}
\delta T_{ij}^{A}=-\eta _{A}%
\begin{pmatrix}
\left( \partial _{1}V_{2}+\partial _{2}V_{1}\right) & \left( -\partial
_{1}V_{1}+\partial _{2}V_{2}\right) \\ 
\left( -\partial _{1}V_{1}+\partial _{2}V_{2}\right) & -\left( \partial
_{1}V_{2}+\partial _{2}V_{1}\right)%
\end{pmatrix}%
+\zeta _{A}\delta _{ij}\left( \partial _{1}V_{2}-\partial _{2}V_{1}\right) .
\end{equation}%
The curl viscosity can also arise from taking $\xi _{ijkl}\propto \delta
_{ij}\epsilon _{kl}$. The curl structure naturally reminds us vortices.\ It
is interesting that the bulk and curl viscosities are associated with the
divergence and the curl of the velocity. Both of them can only exist in
systems without scaling invariance due to its trace like structure in the
energy momentum tensor. It is easy to generalize the above discussion to
higher dimensions. However, the Hall and curl viscosities can only exist in
two dimensions as depicted in Figure 1.

\begin{figure}[tbp]
\includegraphics[width=0.45\textwidth]{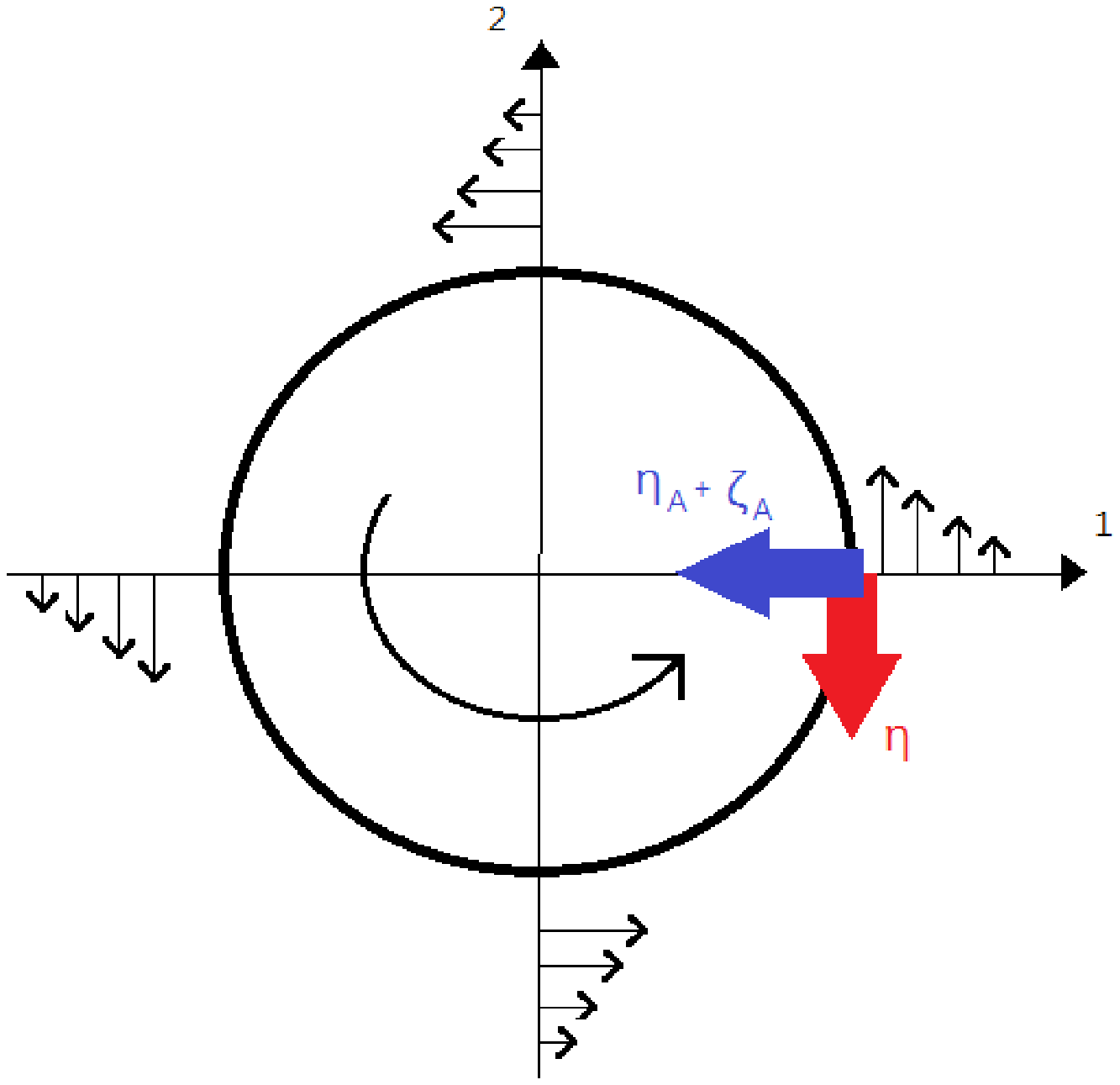} \includegraphics[width=0.45%
\textwidth]{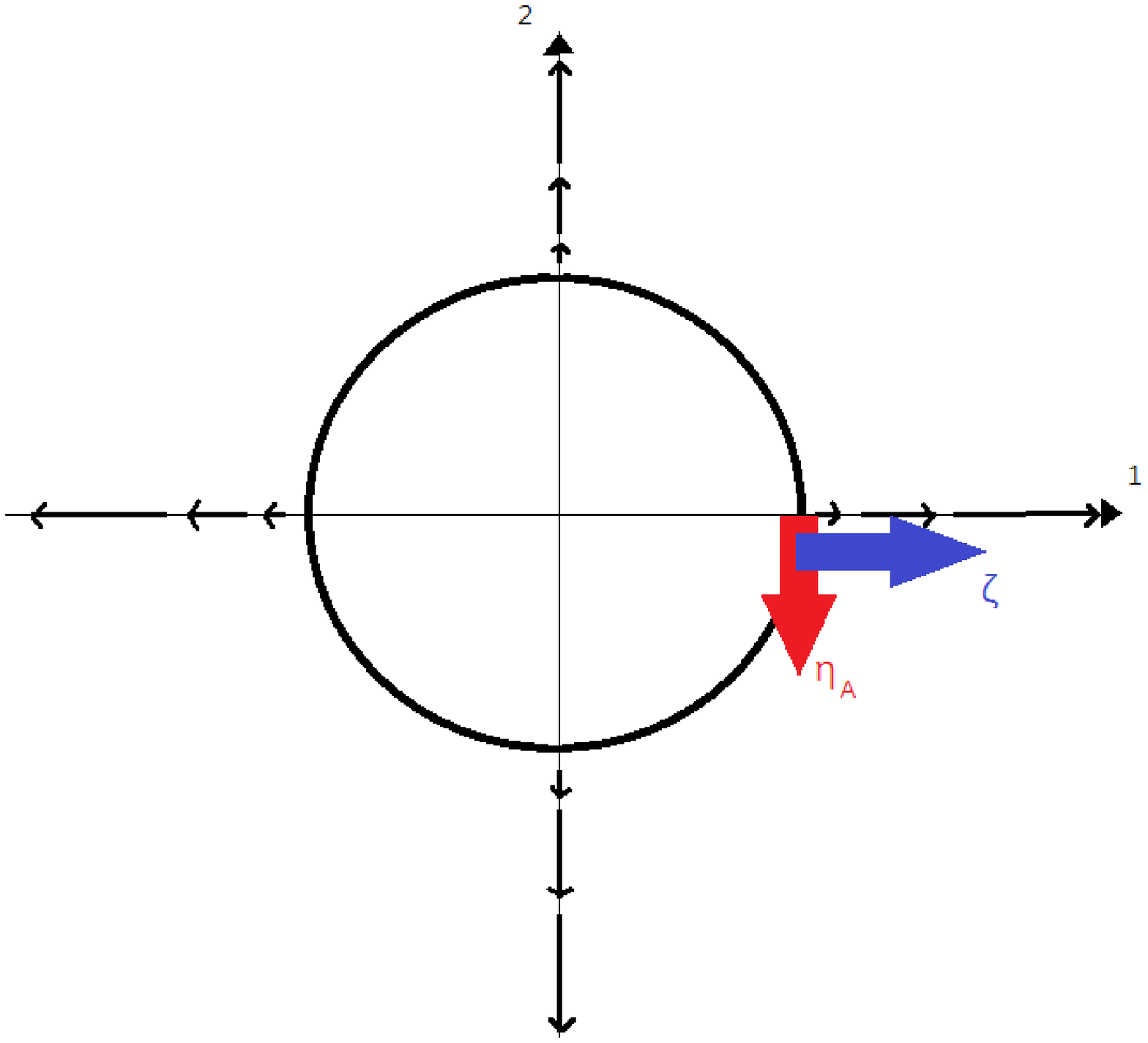}
\caption{Cartoon pictures showing the force directions associated with
various viscosities against non-uniform fluid velocity in the $(x_{1},x_{2})$
plane. The usual bulk viscosity ($\protect\zeta $), the sum of Hall
viscosity ($\protect\eta _{A}$) and curl viscosity ($\protect\zeta _{A}$)
are generated by non-zero curl of velocity, while the shear viscosity ($%
\protect\eta $) and Hall viscosity ($\protect\eta _{A}$) along are generated
by non-zero divergence of velocity. Notice that the force flips its
direction when the direction of the velocity field, which is depicted by the
arrows outside the spheres, is reversed.}
\end{figure}

The Hall and curl viscosities have distinct transformation properties from
the shear and bulk viscosities under parity. Under the coordinate reflection 
$(x_{1},x_{2})\rightarrow (-x_{1},x_{2})$, $\delta T_{ij}\rightarrow \left(
-1\right) ^{i+j}\delta T_{ij}$ while $\delta T_{ij}^{A}\rightarrow \left(
-1\right) ^{i+j+1}\delta T_{ij}^{A}$. Since $\delta T_{ij}$ exists in parity
conserving systems, $\delta T_{ij}^{A}$ only exists in parity violating
systems.

In summary, we need to work in $(2+1)$ dimensional parity violating systems
to study the Hall and curl viscosities. In the following section we will
explicitly construct a holographic model and calculate the Hall viscosity of
the boundary fluid.

\section{The holographic set up}

Following the discussion of the previous section, we will consider a four
dimensional bulk action as the holographic dual to a three dimensional
boundary theory. It is given by a four dimensional Einstein action with a
negative cosmological constant; the matter sector includes an abelian
Yang-Mills $F_{\mu \nu }$ and a pseudo scalar field $\theta $: 
\begin{eqnarray}
\mathcal{L} &=&R-\frac{6}{L^{2}}-\frac{1}{4}F^{2}  \notag  \label{action} \\
&&-\frac{1}{2}(\partial \theta )^{2}-V(\theta )-\frac{\lambda }{4}\theta {}\
^{\ast }RR.
\end{eqnarray}%
The the bulk action conserves parity, so $\theta $ is a pseudo scalar from
the last term on the Lagrangian which is important to introduce parity
violation to the boundary theory through the $\theta $ condensate. The $%
F^{2} $ term is the only difference between our model and Saremi and~Son's.
In our model a charged black hole solution is allowed. We will recover their
result by taking the black hole charge to zero.

In this paper, we will only focus on\textbf{\ }the following form of the
potential, 
\begin{equation}
V(\theta )=\frac{1}{2}m^{2}\theta ^{2}+\frac{1}{4}c\theta ^{4}.
\label{potential}
\end{equation}%
As discussed in \cite{Iqbal:2010eh}, the second term is necessary to have
consistent solution at $T=0$. We will study the probed limit of the scalar
field by sending $\theta \rightarrow \varepsilon \theta $ and $\lambda
\rightarrow \varepsilon \lambda $ for small $\varepsilon $. Thus, at leading
order in $\varepsilon $, we only need to solve for the equation of motion
governed by the upper line of (\ref{action}). Then the background is exactly
a charged black brane in $AdS_{4}$ spacetime and the Hall viscosity $\eta
_{A}$ can be recovered at the $\mathcal{O}(\varepsilon ^{2})$ order, since $%
\eta _{A}\rightarrow \varepsilon ^{2}\eta _{A}$ in this probe limit.

The charged black brane solution is given by the metric: 
\begin{equation}
ds^{2}=2dvdr-r^{2}f(r)dv^{2}+r^{2}(dx^{2}+dy^{2}),
\end{equation}%
where 
\begin{equation}
f(r)=\frac{1}{L^{2}}-\frac{M}{r^{3}}+\frac{Q^{2}}{r^{4}}.
\end{equation}%
and the abelian gauge field\footnote{%
The $A_{r}$ component of gauge potential can be gauged away with no
contribution to the equation of motion. However, we keep it here to show
that under a proper coordinate transformation, $v=t+h(r)$ where $h^{\prime
}(r)=\frac{1}{r^{2}f(r)}$, this black brane solution can be brought to the
usual diagonal coordinate given by $(t,r,x,y)$.}: 
\begin{equation}
A=2\frac {Q}{r_H}(1-\frac{r_{H}}{r})dv-\frac{Q}{r^{3}f}dr.
\end{equation}%
Here black brane mass and electric charge are $M$ and $Q$. The horizon is at 
$r=r_{H}$. The metric is asymptotically $AdS_{4}$ with curvature radius $L$.
It is convenient to work in the units of $L=1$ and rescale the horizon to $%
r_{H}=1$\footnote{%
The action and equations of motion are invariant under the following
scaling: 
\begin{eqnarray}
r &\rightarrow &cr,(v,x,y)\rightarrow c^{-1}(v,x,y),  \notag \\
Q &\rightarrow &c^{2}Q,M\rightarrow c^{3}M,  \notag \\
\theta &\rightarrow &\theta ,A\rightarrow A,f\rightarrow f.  \label{scaling}
\end{eqnarray}%
Here we adopt the convention of \cite{Iqbal:2010eh} where $Q \to Q r_H^2$, $%
r\to r_Hr$ and $(v,x,y) \to r_H^{-1}(v,x,y)$.} 
\begin{equation}
f(r)=1-\frac{1+3\kappa }{r^{3}}+\frac{3\kappa }{r^{4}},\qquad A= 2 \sqrt{3
\kappa} (1- \frac{1}{r})dv.
\end{equation}%
The charged black brane in the bulk corresponds to a boundary field theory
at finite temperature $T$ and chemical potential $\mu $, that is 
\begin{equation}
T=\frac{3}{4\pi }(1-\kappa ),\qquad \mu =2\sqrt{3\kappa }.
\end{equation}%
We remark that $\kappa =0$ corresponds to a neutral black brane with zero
chemical potential and $\kappa =1$ corresponds to an extremal black brane at
zero temperature.

The equation of motion the probed neutral pseudo scalar reads, 
\begin{equation}
\theta ^{\prime \prime }+(\frac{f^{\prime }}{f}+\frac{4}{r})\theta ^{\prime
}-\frac{V^{\prime }(\theta )}{r^{2}f}=0.
\end{equation}%
Near the boundary, the asymptotic behavior of pseudo scalar is 
\begin{equation}
\theta =\frac{J}{r^{\Delta _{-}}}+\frac{\mathcal{O}}{r^{\Delta _{+}}}+\cdots
,
\end{equation}%
with 
\begin{equation}
\Delta _{\pm }=\frac{3}{2}\pm \sqrt{\frac{9}{4}+m^{2}L^{2}}
\end{equation}%
We remark that in our construction, the mode $J$ can be consistently turned
off and $\mathcal{O}$ is identified as the condensate in the boundary 
\footnote{%
For $-9/4<m^{2}L^{2}<-5/4$, $J$ and $\mathcal{O}$ are both renormalizable
and one is free to choose either one as source and the other as condensate%
\cite{Klebanov:1999tb}.}. However, this was not possible in the original
construction with neutral black brane \cite{Saremi:2011ab} where $J$ can be
turned off only if $c<-\frac{3}{4}$ \cite{Hertog:2006rr} which violates the
positive energy theorem and hence it is not a stable solution. In our model,
the $\theta ^{4}$ term is required to make the $\theta $ solution regular at
the horizon \cite{Iqbal:2010eh}.

\begin{figure}[tbp]
\includegraphics[width=0.6\textwidth]{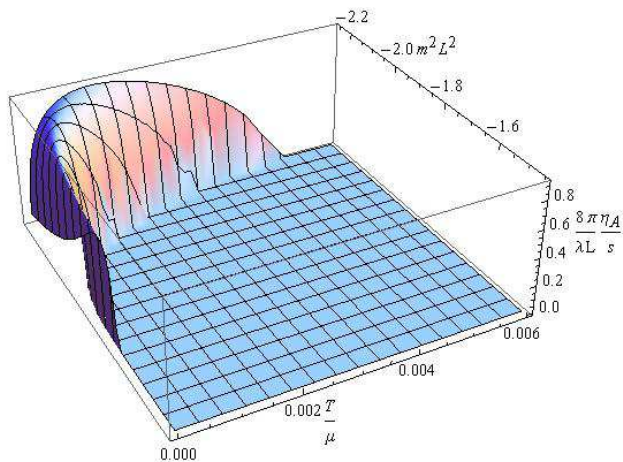}
\caption{$\protect\eta _{A}/s\protect\lambda L$ as a function of $T/\protect%
\mu $ and $m^{2}L^{2}$.}
\end{figure}

\section{The Hall viscosity}

The detail derivation of the viscosities is presented in the appendix. We
have first included the back reaction of $\theta $ as was done in \cite%
{Saremi:2011ab}, then take the probe limit ($\theta \rightarrow \varepsilon
\theta $ and $\lambda \rightarrow \varepsilon \lambda $) to the final result
of the viscosity expression. The expression for $\eta _{A}$, which appears
at ${\mathcal{O}}(\varepsilon ^{2})$, is identical to that obtained in \cite%
{Saremi:2011ab}.

From the derivation in the appendix, we obtain the shear viscosity of the
universal value as expected: 
\begin{equation}
\frac{\eta }{s}=\frac{1}{4\pi },
\end{equation}%
where $s$ is the entropy density. Theses combinations are dimensionless and
are invariant under the scaling of Eq.(\ref{scaling}).

The Hall viscosity in our charged black brane background takes the same form
as the case of the neutral black brane background \cite{Saremi:2011ab}, that
is 
\begin{equation}
\eta _{A}=-\frac{1}{8\pi G_{N}}\frac{\lambda }{4}\left. \frac{r^{4}f^{\prime
}(r)\theta ^{\prime }(r)}{H(r)^{2}}\right\vert _{r=r_{H}}.  \label{etaA}
\end{equation}%
The dimensionless and scale invariant combination yields 
\begin{equation}
\frac{\eta _{A}}{s}=-\frac{\lambda }{8\pi }\left. \frac{r^{4}f^{\prime
}(r)\theta ^{\prime }(r)}{H(r)^{2}}\right\vert _{r=r_{H}}.  \label{etaAs}
\end{equation}

\begin{figure}[tbp]
\includegraphics[width=0.6\textwidth]{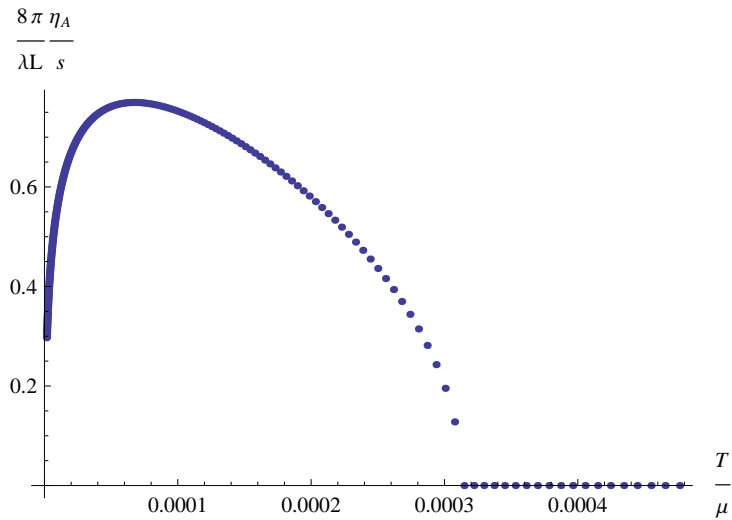}
\caption{$\protect\eta _{A}/s\protect\lambda L$\textbf{\ }vs. $T/\protect\mu 
$ for $m^{2}L^{2}=-2.$}
\end{figure}

In Eq.(\ref{etaAs}), $\eta _{A}/s$ vanishes when the solution of $\theta $
is trivial ($\theta (r)=0$), which happened in the symmetric phase, or when $%
\theta $ is a constant field. In the former case, parity is not broken in
the bulk. Then by the correspondence, it will not be broken at the boundary
either. Likewise, in the latter case, when $\theta $ is a constant, the $%
^{\ast }RR$ term is just a surface term in the action which has no effect to
the bulk equations of motion. Hence it does not contribute to $\eta _{A}$
either. Therefore, it should not be a surprise that the phase diagram for $%
\eta _{A}/s$ is very similar to that with the neutral scalar hair of Ref. 
\cite{Iqbal:2010eh} with just one difference---$\eta _{A}/s$ vanishes when $%
T=0$. This comes from the factor $f^{\prime }(r_H)\propto T$. One peculiar
feature of this model is that the entropy of the charged black hole does not
vanish at zero temperature. Perhaps in models with zero entropy at zero
temperature, $\eta _{A}/s$ stays finite at zero temperature. We show $\eta
_{A}/s\lambda L$ as a function of $T/\mu $ and $m^{2}L^{2}$ in Fig. 2. These
three quantities are all scale invariant and dimensionless.

In Fig. 3, the dependence of the scale invariant, dimensionless quantities $%
\eta _{A}/s\lambda L$\textbf{\ }and $T/\mu $ is shown for $m^{2}L^{2}=-2$. $%
\eta _{A}/s\propto T$ as $T\rightarrow 0$\ due to $f^{\prime }(r_{H})\propto
T$ in Eq.(\ref{etaA}). When $T\rightarrow T_{c}$, $\eta _{A}/s$ vanishes.
The analytic approximation performed in the Appendix suggests that critical
exponent is of mean field value: $\eta _{A}/s\propto (1-T/T_{c})^{1/2}$ as $%
T\rightarrow T_{c}$. One can also see that $\eta _{A}\rightarrow 0$ as we
take $\mu \rightarrow 0$ and the black hole becomes charge neutral.

In Fig. 4, $\eta _{A}/s\lambda L$\textbf{\ }vs. $m^{2}L^{2}$ is plotted for $%
T/\mu =7.55\times 10^{-6}$. The critical $m^{2}L^{2}$ is smaller than the
critical value $m^{2}L^{2}=-1.5$ at zero $T$ because it is harder to form
the condensate at higher $T$.

In our model, the non-zero Hall viscosity arises because parity is broken
spontaneously. The non-zero classical solution (or equivalently, vacuum
expectation value) of $\theta $ yields a pseudo scalar condensate at the
boundary which is a necessary condition to have non-zero Hall viscosity.

\begin{figure}[tbp]
\includegraphics[width=0.6\textwidth]{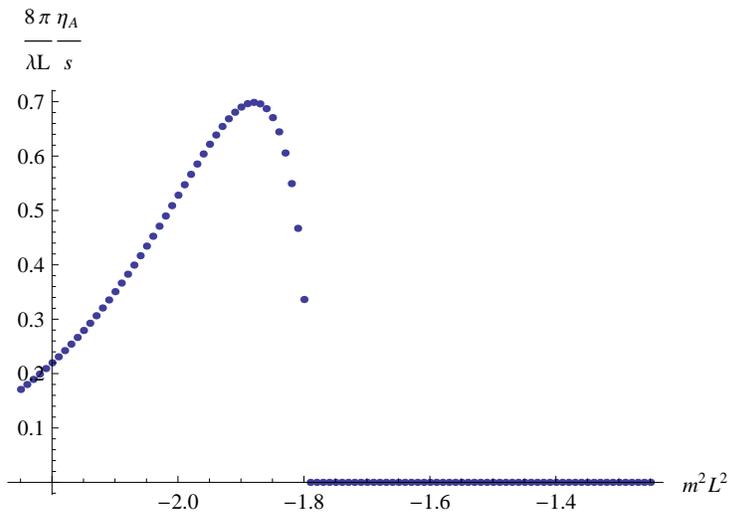}
\caption{$\protect\eta _{A}/s\protect\lambda L$\textbf{\ }vs. $m^{2}L^{2}$
for $T/\protect\mu =7.55\times 10^{-6}.$}
\end{figure}

\section{Conclusion}

We have modified the holographic model of Saremi and~Son \cite{Saremi:2011ab}
by using a charged black brane, instead of a neutral one, such that when the
bulk pseudo scalar ($\theta $) potential is made of $\theta ^{2}$ and $%
\theta ^{4}$ terms, parity can still be broken spontaneously in the boundary
theory. In our model, the 3+1 dimensional bulk has a pseudo scalar coupled
to the gravitational Chern-Simons term in the anti de Sitter charged black
brane back ground. Parity could be broken spontaneously in the bulk by the
pseudo scalar hairy solution and give rise to non-zero Hall viscosity at the
boundary theory.

This study does not exclude a non-vanishing Hall viscosity in Saremi
and~Son's model be found with a more general potential. It is interesting to
investigate the Hall viscosity in other parity-broken holographic condensed
matter systems, such as the D-wave superconductors\cite%
{Chen:2010mk,Chen:2011ny}. We will report it in a future project.

\section{Appendix}

\subsection{Derivation of Hall viscosity}

Here we detail the Hall viscosity derivation with the charged black brane
solution. The hydrodynamics of charged fluid has been extensively studied in
the holographic set up \cite{nabamita}. The general procedure to calculate
the holographic hydrodynamic transport coefficients has been given in \cite%
{minwalla}. We largely follow the procedures adopted in \cite%
{minwalla,Saremi:2011ab} with the neutral black brane solution. The
equations of motion by varying the action (\ref{action}) with respect to the
metric, the scalar and the gauge field are as

\begin{eqnarray}
&&R_{MN}-\frac{1}{2}g_{MN}R+\Lambda g_{MN}-\lambda C_{MN}=T_{MN}(\theta
)+T_{MN}(A),  \notag \\
&&\nabla ^{2}\theta =\frac{dV}{d\theta }+\frac{\lambda }{4}^{\ast }RR, 
\notag \\
&&\nabla _{M}F^{MN}=0,  \label{EOM}
\end{eqnarray}%
where%
\begin{eqnarray}
&&T_{MN}(\theta )=\frac{1}{2}\partial _{M}\theta \partial _{N}\theta -\frac{1%
}{4}g_{MN}(\partial \theta )^{2}-\frac{1}{2}g_{MN}V(\theta ),  \notag \\
&&T_{MN}(A)=\frac{1}{2}F_{M}^{\text{ \ \ \ }A}F_{NA}-\frac{1}{8}%
g_{MN}F_{AB}F^{AB},  \notag
\end{eqnarray}%
and $C_{MN}$ is called Cotton tensor coming from the gravitational
Chern-Simons term 
\begin{equation}
C^{MN}=\frac{1}{2}\left[ \partial _{A}\theta (\epsilon ^{AMBC}\nabla
_{B}R_{C}^{N}+\epsilon ^{ANBC}\nabla _{B}R_{C}^{M})+\nabla _{A}\partial
_{B}\theta (^{\ast }R^{AMBN}+^{\ast }R^{ANBM})\right] ,  \notag
\end{equation}%
where $\epsilon ^{AMBC}$ is the usual four dimensional Levi-Civita tensor.

An ansatz satisfying the equations of motion is

\begin{eqnarray}
&&ds^{2}=-2H(r,b,q)u_{\mu }dx^{\mu }dr-r^{2}f(r,b,q)u_{\mu }u_{\nu }dx^{\mu
}dx^{\nu }+r^{2}P_{\mu \nu }dx^{\mu }dx^{\nu },  \notag \\
&&\theta =\theta (r,b,q),  \notag \\
&&A=A(r,b,q)u_{\mu }dx^{\mu }.
\end{eqnarray}%
This ansatz describes a boosted black brane solution alone the boundary
coordinates. Then following the standard procedure of the Fluid/Gravity
correspondence, we perturb the system away from equilibrium by promoting the
velocity $u^{\mu }$, mass $b$ and charge $q$ to vary slowly with respect to
the boundary coordinates. In the co-moving frame where the fluid
two-velocity is zero at the origin of the boundary coordinates ($x^{\mu }=0$%
), we Taylor expand quantities near the origin to the first derivative
order: 
\begin{eqnarray}
&&u^{\mu }=(1,x^{\mu }\partial _{\mu }\beta ^{i}),  \notag \\
&&b=b_{0}+x^{\mu }\partial _{\mu }b,  \notag \\
&&q=q_{0}+x^{\mu }\partial _{\mu }q,  \notag \\
&&f(r,b,q)=f(r)+\frac{\partial f}{\partial b}x^{\mu }\partial _{\mu }b+\frac{%
\partial f}{\partial q}x^{\mu }\partial _{\mu }q=f(r)+\delta f,  \notag \\
&&H(r,b,q)=H(r)+\frac{\partial H}{\partial b}x^{\mu }\partial _{\mu }b+\frac{%
\partial H}{\partial q}x^{\mu }\partial _{\mu }q=H(r)+\delta H,  \notag \\
&&A(r,b,q)=A(r)+\frac{\partial A}{\partial b}x^{\mu }\partial _{\mu }b+\frac{%
\partial A}{\partial q}x^{\mu }\partial _{\mu }q=A(r)+\delta A,  \notag \\
&&\theta (r,b,q)=\theta (r)+\frac{\partial \theta }{\partial b}x^{\mu
}\partial _{\mu }b+\frac{\partial \theta }{\partial q}x^{\mu }\partial _{\mu
}q=\theta (r)+\delta \theta .
\end{eqnarray}

\bigskip Substitute these into the ansatz, we get

\begin{eqnarray}
ds^{2} &=&2H(r)dvdr-r^{2}f(r)dv^{2}+r^{2}dx^{i}dx^{i} \\
&&+\epsilon \left[ 2\delta Hdvdr-r^{2}\delta fdv^{2}-2H(r)x^{\mu }\partial
_{\mu }\beta ^{i}dx^{i}dr-2r^{2}(1-f(r))x^{\mu }\partial _{\mu }\beta
^{i}dx^{i}dv\right] ,  \notag \\
\theta &=&\theta (r)+\epsilon \delta \theta ,  \notag \\
A &=&-A(r)dv+\epsilon (-\delta Adv+A(r)x^{\mu }\partial _{\mu }\beta
^{i}dx^{i}),
\end{eqnarray}%
where we have added the parameter $\epsilon $ to keep track of how many
derivatives on the boundary coordinates each term has.

Note that after we promote the parameter to be dependent on the boundary
coordinates, the ansatz no longer satisfies the equations of motion. Hence
we add corrections order by order to the metric, scalar and gauge fields
such that, order by order, the whole metric, scalar and gauge fields still
satisfy the equations of motion. To calculate the Hall viscosity, it
suffices to consider the symmetric traceless part of the correction to the
metric:%
\begin{eqnarray}
&&ds^{2}=\epsilon \left( \frac{k(r)}{r^{2}}%
dv^{2}+2h(r)dvdr-r^{2}h(r)dx^{i}dx^{i}+\frac{2}{r}a^{i}(r)dvdx^{i}+r^{2}%
\alpha _{ij}(r)dx^{i}dx^{j}\right) ,  \notag \\
&&\theta =\epsilon \theta _{cor},  \notag \\
&&A=\epsilon (A_{cor}^{v}(r)dv+A_{cor}^{x}(r)dx+A_{cor}^{y}(r)dy).
\end{eqnarray}

In this case, the trace-reversed form of the Einstein equations is more
convenient, which is given by 
\begin{equation}
E_{MN}=R_{MN}-\Lambda g_{MN}-\lambda C_{MN}-d_{MN}=0,
\end{equation}%
where 
\begin{eqnarray}
&&d_{MN}=d_{MN}(\theta )+d_{MN}(A)=T_{MN}-\frac{1}{2}g_{MN}T,  \notag \\
&&d_{MN}(\theta )=\frac{1}{2}\partial _{M}\theta \partial _{N}\theta +\frac{1%
}{2}g_{MN}V(\theta ),  \notag \\
&&d_{MN}(A)=-\frac{1}{2}F_{M}^{\text{ \ \ \ }A}F_{AN}-\frac{1}{8}%
g_{MN}F_{AB}F^{AB}
\end{eqnarray}

Substitute all into the Einstein equations and collect the first order term
from the xy-component of the trace-reversed Einstein equation, we get 
\begin{eqnarray}
&&\frac{1}{H}\frac{d}{dr}\left[ -\frac{1}{2}\frac{r^{4}f}{H}\frac{d}{dr}%
\alpha _{xy}\right] +\left[ \frac{r^{3}H^{\prime }f}{H^{3}}-\frac{%
r^{3}f^{\prime }}{H^{2}}-\frac{3r^{2}f}{H^{2}}+3r^{2}-\frac{r^{2}}{2}%
V(\theta )-\frac{r^{2}A^{\prime 2}}{4H^{2}}\right] \alpha _{xy}  \notag \\
&=&\frac{r}{H}(\partial _{x}\beta _{y}+\partial _{y}\beta _{x})+\frac{%
\lambda }{4H}\frac{d}{dr}(\frac{r^{4}f^{\prime }\theta ^{\prime }}{H^{2}}%
)(\partial _{x}\beta _{x}-\partial _{y}\beta _{y})
\end{eqnarray}%
However, the zeroth order of the xx-component of the trace-reversed Einstein
equation yields 
\begin{equation}
\frac{r^{3}H^{\prime }f}{H^{3}}-\frac{r^{3}f^{\prime }}{H^{2}}-\frac{3r^{2}f%
}{H^{2}}+3r^{2}-\frac{r^{2}}{2}V(\theta )-\frac{r^{2}A^{\prime 2}}{4H^{2}}=0.
\end{equation}%
Therefore we obtain a differential equation for $\alpha _{xy}$, 
\begin{equation}
\frac{d}{dr}\left[ -\frac{1}{2}\frac{r^{4}f}{H}\frac{d}{dr}\alpha _{xy}%
\right] =r(\partial _{x}\beta _{y}+\partial _{y}\beta _{x})+\frac{\lambda }{4%
}\frac{d}{dr}(\frac{r^{4}f^{\prime }\theta ^{\prime }}{H^{2}})(\partial
_{x}\beta _{x}-\partial _{y}\beta _{y}).
\end{equation}%
And hence%
\begin{equation}
\alpha _{xy}(r)=\int_{r}^{\infty }\frac{2H(t)dt}{t^{4}f(t)}\int_{r_{H}}^{t}dz%
\left[ z(\partial _{x}\beta _{y}+\partial _{y}\beta _{x})+\frac{\lambda }{4}%
\frac{d}{dz}(\frac{z^{4}f^{\prime }\theta ^{\prime }}{H^{2}})(\partial
_{x}\beta _{x}-\partial _{y}\beta _{y})\right] .
\end{equation}%
As in \cite{Saremi:2011ab}, we use the following formula to compute the
asymptotic form, 
\begin{equation}
r^{n}\alpha _{xy}(r)\rightarrow -\frac{r^{n+1}}{n}\frac{d}{dr}\alpha _{xy}(r)%
\text{ \ as }r\rightarrow \infty .
\end{equation}%
And from \cite{de Haro:2000xn}, the boundary energy momentum tensor for odd
boundary dimension is given by%
\begin{equation}
\langle T_{ij}\rangle =\frac{d}{16\pi G_{N}}g_{(d)ij},
\end{equation}%
where%
\begin{equation}
g(x^{\mu },r)=g_{(0)}+\frac{1}{r^{2}}g_{(2)}+\cdots +\frac{1}{r^{d}}%
g_{(d)}+\cdots .
\end{equation}%
Therefore, all we have to do is to find the constant part of $r^{3}\alpha
_{xy}$. Hence%
\begin{eqnarray*}
r^{3}\alpha _{xy}(r) &\rightarrow &-\frac{r^{4}}{3}\frac{d}{dr}\alpha
_{xy}(r) \\
&=&-\frac{r^{4}}{3}\frac{d}{dr}\int_{r}^{\infty }\frac{2H(t)dt}{t^{4}f(t)}%
\int_{r_{H}}^{t}dz\left[ z(\partial _{x}\beta _{y}+\partial _{y}\beta _{x})+%
\frac{\lambda }{4}\frac{d}{dz}(\frac{r^{4}f^{\prime }\theta ^{\prime }}{H^{2}%
})(\partial _{x}\beta _{x}-\partial _{y}\beta _{y})\right] \\
&=&\frac{r^{4}}{3}\frac{2H(r)}{r^{4}f(r)}\int_{r_{H}}^{r}dz\left[ z(\partial
_{x}\beta _{y}+\partial _{y}\beta _{x})+\frac{\lambda }{4}\frac{d}{dz}(\frac{%
r^{4}f^{\prime }\theta ^{\prime }}{H^{2}})(\partial _{x}\beta _{x}-\partial
_{y}\beta _{y})\right] \\
&=&\frac{2H}{3f}\left[ \frac{z^{2}}{2}(\partial _{x}\beta _{y}+\partial
_{y}\beta _{x})+\frac{\lambda }{4}(\frac{r^{4}f^{\prime }\theta ^{\prime }}{%
H^{2}})(\partial _{x}\beta _{x}-\partial _{y}\beta _{y})\right] _{rH}^{r}
\end{eqnarray*}

Since $f$ and $H$ asymptotically approach 1 and we shift the horizon to $%
r_{H}=1$, we get the xy-component of the boundary energy momentum tensor as%
\begin{eqnarray*}
\langle T_{xy}\rangle &=&\frac{3}{16\pi G_{N}}g_{(3)xy} \\
&=&-\frac{1}{16\pi G_{N}}(\partial _{x}\beta _{y}+\partial _{y}\beta _{x})-%
\frac{1}{8\pi G_{N}}\left[ \frac{\lambda }{4}(\frac{r^{4}f^{\prime }\theta
^{\prime }}{H^{2}})(\partial _{x}\beta _{x}-\partial _{y}\beta _{y})\right]
_{r=r_{H}}
\end{eqnarray*}%
The first term is the usual shear mode with 
\begin{equation}
\eta =\frac{1}{16\pi G_{N}}
\end{equation}%
which recovers%
\begin{equation}
\frac{\eta }{s}=\frac{1}{4\pi }.
\end{equation}%
The second term is proportional to the Hall viscosity which yields%
\begin{equation}
\eta _{A}=-\frac{1}{8\pi G_{N}}\frac{\lambda }{4}\left. \frac{r^{4}f^{\prime
}(r)\theta ^{\prime }(r)}{H(r)^{2}}\right\vert _{r=r_{H}}.
\end{equation}%
The dimensionless combination 
\begin{equation}
\frac{\eta _{A}}{s}=-\frac{\lambda }{8\pi }\left. \frac{r^{4}f^{\prime
}(r)\theta ^{\prime }(r)}{H(r)^{2}}\right\vert _{r=r_{H}}
\end{equation}%
is independent of the scaling.

\subsection{Analytic Approximation}

Here we will apply an approximation to obtain an analytic expression for
Hall viscosity in terms of the condensate in the boundary theory. We use the
new coordinate $z=1/r$ for convenience. Near the horizon $z=1$, we can
expand $\theta (z)$ as 
\begin{equation}
\theta (z)\simeq \theta (1)-\frac{m^{2}}{3(1-\kappa )}\theta (1)(1-z).
\end{equation}%
On the other hand, near the boundary $z=0$, one has 
\begin{equation}
\theta (z)\simeq \mathcal{O}z^{\Delta _{+}}.
\end{equation}%
By matching above expressions in the middle $z=1/2$, one can identify
approximately 
\begin{equation}
\theta (1)\simeq \frac{\mathcal{O}}{1-\frac{m^{2}}{6(1-\kappa )}}2^{-\Delta
_{+}}.
\end{equation}%
Therefore one can express the Hall viscosity as a function of temperature
and the condensate $\mathcal{O}$: 
\begin{equation}
\eta _{A}=-\frac{\lambda }{32\pi G_{4}}\frac{m^{2}\mathcal{O}}{1-\frac{m^{2}%
}{8\pi T}}2^{-\Delta _{+}}.
\end{equation}%
Near the critical $T_{c}$, the condensate has the mean field $T$ dependence
for a second order phase transition \cite{Zeng:2010zn}: 
\begin{equation}
\mathcal{O}\propto T_{c}^{\Delta _{+}}(1-\frac{T}{T_{c}})^{1/2}\theta
(T_{c}-T),
\end{equation}%
hence near $T_{c}$ one has 
\begin{equation}
\eta _{A}\propto (1-\frac{T}{T_{c}})^{1/2}\theta (T_{c}-T).
\end{equation}

\begin{acknowledgments}
This work is supported in part by the National Science Council, National
Center for Theoretical Science and the CASTS of NTU.
\end{acknowledgments}


\end{document}